\documentclass{llncs}
\usepackage{graphicx}
\usepackage{subcaption}
\usepackage{enumitem}
\usepackage{color}
\usepackage{algorithm}
\usepackage{algorithmicx}
\usepackage[noend]{algpseudocode}
\usepackage{amsmath}

\usepackage{tabularx}                      
\usepackage{booktabs}                  

\graphicspath{ {./Images/} }

\newcommand{\red}[1]{{\color{red}#1}} 
\newcommand{\qmarks}[1]{``#1''}

\begin{document}

\title{Experiments and a User Study for Hierarchical Drawings of Graphs}




%
\titlerunning{Experiments and a User Study for Hierarchical Drawings of Graphs}
%
\author{P. Lionakis \and G. Kritikakis \and I. G. Tollis}
%
%
\institute{Computer Science Department, University of Crete, GREECE \email{{\{lionakis,gkrit,tollis\}@csd.uoc.gr}}
\\}
%
\maketitle              
\begin{abstract}
We present  experimental results and a user study for
hierarchical drawings of graphs.
A detailed hierarchical graph drawing technique that is based on the
\textit{Path Based Framework
(PBF)} is presented.  Extensive edge bundling is applied to draw all
edges of the graph and
the height of the drawing is minimized using compaction. 
The drawings produced by this framework are compared to drawings produced by the
well known Sugiyama framework in terms of area, number of bends, number of crossings, and execution time.  
The new algorithm runs very
fast and produces drawings that are readable and efficient.
Since there are advantages (and disadvantages) to both frameworks, we performed a user study and the results show that the drawings produced by the new framework are well received in terms of clarity, readability, and usability. 
Hence, the new technique offers an interesting alternative to drawing hierarchical graphs, and is
especially useful in applications where user defined paths are important and need to be highlighted.

\end{abstract}
\section{Introduction}

Hierarchical graphs are very important for many applications in several areas of research and business because they often represent hierarchical relationships between objects in a structure. They are directed (often acyclic) graphs and their visualization has received significant attention recently~\cite{DBLP:books/ph/BattistaETT99,kw,handbook}. An experimental study of four algorithms specifically designed for (Directed Acyclic Graphs) DAGs was presented in~\cite{DBLP:conf/gd/BattistaGLPTTVV96}. DAGs are often used to describe processes containing some long paths, such as in PERT applications see for example~\cite{di1989automatic,fisher1983stochastic}. The paths can be either application based, e.g. critical paths, user defined, or automatically generated paths.
A new framework  to visualize directed graphs and their hierarchies was introduced in~\cite{ortali2018algorithms,JGAA-502}.  It is based on a path/channel decomposition of the graph and is called \textit{(Path-Based Framework or PBF)}. It computes readable hierarchical visualizations in two phases by \qmarks{hiding} (\emph{abstracting}) some selected edges, while maintaining the complete reachability information of a graph. However, these drawings are not satisfactory to users that need to visualize the whole graph.
\par
In this paper we extend the hierarchical graph drawing framework of~\cite{ortali2018algorithms,JGAA-502} in two directions: a) we draw all edges of the graph and use extensive edge bundling, and b) we minimize the height of the drawing using a compaction technique. To reduce the width we apply algorithms similar to task scheduling.  The total time required for these extensions is $O(m + n \log n)$, where $m$ is the number of edges and $n$ the number of nodes of a graph $G$. In this framework, the edges of $G$ are naturally split into three categories: \textit{path edges, path transitive edges, and cross edges}. Path edges connect consecutive vertices in the same path. Path transitive edges connect non-consecutive vertices in the same path. Cross edges connect vertices that belong to different paths.

\par
The path-based framework departs from the typical Sugiyama Framework~\cite{DBLP:journals/tsmc/SugiyamaTT81} and it consists of two phases: (a) Cycle Removal, (b) the path/channel decomposition and hierarchical drawing step.  It is based on the idea of partitioning the vertices of a graph into node disjoint paths/channels, where in a channel consecutive nodes are connected by a path but not necessarily connected by an edge. In the rest, we only use the term ``path" but of course our algorithms work also for ``channels."  The vertices in each path are drawn vertically aligned on some x-coordinate; next the edges between vertices that belong to different paths are drawn. Note that there are several algorithms that compute a path decomposition of minimum cardinality in polynomial time~\cite{DBLP:journals/siamcomp/HopcroftK73},\cite{DBLP:conf/recomb/KuosmanenPGCTM18},\cite{DBLP:conf/stoc/Orlin13},\cite{DBLP:journals/siamcomp/Schnorr78}.

The Sugiyama Framework has been extensively used in practice, as manifested by the fact that various systems are using it to implement hierarchical drawing techniques.  
The comparative study of~\cite{DBLP:conf/gd/BattistaGLPTTVV96} concluded that the Sugiyama-style algorithms performed better in most of the metrics.  For more recent information regarding this framework see~\cite{handbook}. Commercial software such as the Tom Sawyer Software TS Perspectives \cite{Tom} and yWorks \cite{yWorks}, use this framework in order to offer automatic visualizations of directed graphs.
Even though it is very popular, the Sugiyama Framework has several limitations: as discussed bellow, most problems and subproblems that are used to optimize the results in various steps of each phase have turned out to be NP-hard.
The overall time complexity of this framework (depending upon implementation) can be as high as $O((nm)^2)$, or even higher if one chooses algorithms that require exponential time.  Another important limitation of this framework is the fact that heuristic solutions and decisions that are made during previous phases (e.g., crossing reduction) will influence severely the results obtained in later phases. Nevertheless, previous decisions cannot be changed in order to obtain better results.

By contrast, in the framework of~\cite{JGAA-502} most problems of the second phase can be solved in polynomial time. If a path decomposition contains $k$ paths, the number of bends introduced is at most $O(kn)$ and the required area is at most $O(kn)$.
Edges between non consecutive vertices in a path (the \emph{path transitive edges}), are not drawn in the framework of~\cite{JGAA-502}. Hence, users that need to visualize all the edges of a given graph are not satisfied by these drawings.

\par
We present experimental results comparing drawings obtained by the extended-PBF to the drawings obtained by running the hierarchical drawing module of OGDF~\cite{DBLP:reference/crc/ChimaniGJKKM13}, which is based on the Sugiyama Framework, and is a quite active research software that implements this framework.  The results show that PBF runs much faster, has better area and less bends, but OGDF has less crossings, especially for sparse graphs. 
\\
\indent
Since the usual metrics like bends, area, and crossings did not lead to concrete conclusions and the two frameworks produce vastly different drawings, we decided to perform a user study
between these two drawing frameworks, in order to obtain feedback from users using these drawings.
The users had to perform a set of tasks on the drawings of some DAGs.  The tasks include determining if two given vertices are connected, finding the length of a shortest path, 
and determining if some vertices are successors of a given vertex.  The users' answers were correct above 90\% for PBF and above 84\% for the Sugiyama Framework (OGDF). The users were also asked to express their preference, in terms of clarity and readability, between the two frameworks.  58.3\% of the users showed a clear preference to using  drawings produced by PBF. Hence, this technique offers an interesting alternative to drawing hierarchical graphs, especially if there are user defined paths that need to be clearly visualized. Section 4 describes a detailed analysis of the user study and the experimental results.

\section{Overview of the Two Frameworks}
\begin{figure}[]
\begin{subfigure}{0.5\textwidth}
  \centering
  \includegraphics[height=13cm ,width=0.7\linewidth]{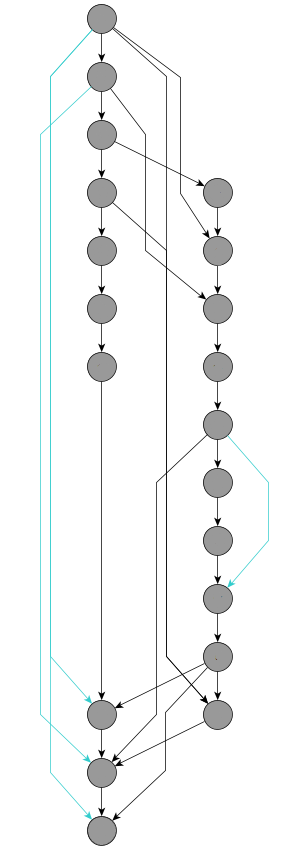} 
  \label{fig:sfig1}
\end{subfigure}%
\begin{subfigure}{.5\textwidth}
  \centering
  \includegraphics[height=13cm ,width=0.65\linewidth]{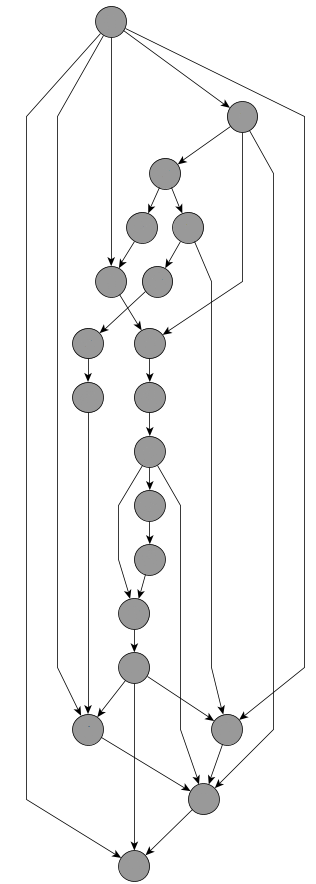} 
  \label{fig:sfig2}
\end{subfigure}
\caption{Example of a DAG $G$ drawn by our proposed framework (left). Same DAG drawn by the Sugiyama framework as implemented in OGDF (right).}
\label{teaser}
\end{figure}

\textbf{}
\label{se:overview}
The two hierarchical drawings shown in Figure~\ref{teaser} demonstrate the significant differences between the two frameworks: 
Part~(a) shows a drawing of $G$ computed by our algorithms that customize PBF. 
Part~(b) shows the drawing of $G$ computed by OGDF. The graph consists of 20 nodes and 31 edges. The drawing computed by our algorithms has 12 crossings, 18 bends, width 10, height 15, and area 150. On the other hand, OGDF computes a drawing that has 5 crossings, 22 bends, width 18, height 15 and area 270. 
Clearly, the two frameworks produce vastly different drawings with their own advantages and disadvantages.


The original Path Based Hierarchical Drawing Framework follows an approach to visualize directed acyclic graphs that hides some edges and focuses on maintaining their reachability information~\cite{JGAA-502}. This framework is based on the idea of partitioning the vertices of the graph $G$ into (a minimum number of) \emph{channels/paths}, that we call \emph{channel/path decomposition} of $G$, which can be computed in polynomial time. Therefore, it is orthogonal to the Sugiyama framework in the sense that it is a vertical decomposition of $G$ into (vertical) paths/channels.  Thus, most resulting problems are \emph{vertically contained}, which makes them simpler, and reduces their time complexity. This framework does not need to introduce any dummy vertices and keeps the vertices of a path \emph{vertically aligned}, which is important for specific applications (such as visualizing critical paths in PERT diagrams~\cite{di1989automatic}).  By contrast, the Sugiyama framework performs a horizontal decomposition of a graph, even though the final result is a vertical (hierarchical) visualization. 
Let  $S_p= \{P_1,...,P_k\}$ be a path decomposition of $G$ such that every vertex $v\in V$ belongs to exactly one of the paths of $S_p$. Any path decomposition naturally splits the edges of $G$ into: (a) \emph{path edges} 
(b) \emph{cross edges} 
and (c) \emph{path transitive edges}.
Given any $S_p$ the main algorithm of~\cite{JGAA-502}, draws the vertices of each path $P_i$ \emph{vertically aligned} on some $x$-coordinate depending on the order of path $P_i$. 
The $y$-coordinate of each vertex is equal to its order in any topological sorting of $G$.
Hence the height of the resulting drawing is $n-1$.  

It also important to highlight that the Path-Based Framework works for any given path decomposition.  Therefore, it can be used in order to draw graphs with user-defined or application-defined paths, as is the case in many applications, see for example~\cite{di1989automatic,fisher1983stochastic}.  If one desires automatically generated paths, there are several algorithms that compute a path decomposition of minimum cardinality in polynomial time~\cite{DBLP:journals/siamcomp/HopcroftK73,DBLP:conf/recomb/KuosmanenPGCTM18,DBLP:conf/stoc/Orlin13,DBLP:journals/siamcomp/Schnorr78}.
Since certain critical paths are important for many applications, it is extremely important to produce clear drawings where all such paths are vertically aligned, see~\cite{JGAA-502}. 
For the rest of this paper, we assume that a path decomposition of $G$ is given as part of the input to the algorithm.

OGDF is a self-contained C++ library of graph algorithms, in particular for (but not restricted to) automatic graph drawing.
The hierarchical drawing implementation of the Sugiyama Framework in OGDF is implemented following~\cite{gansner1993technique,sander1996layout} and it uses the following default choices: For the first phase of Sugiyama, it uses the $LongestPathRanking$ (to assign vertices into layers) which implements the well-known longest-path ranking algorithm.  Next, it performs crossing minimization by using the $BarycenterHeuristic$. This module performs two-layer crossing minimization and is applied during the top-down and bottom-up traversals~\cite{DBLP:reference/crc/ChimaniGJKKM13}.  The crossing minimization is repeated 15 times, and it keeps the best. 
Finally, the final coordinates (drawing) are computed with $FastHierarchyLayout$ layout of OGDF.

\section{Computing Compact and Bundled Drawings}

We present an extension of the framework of~\cite{JGAA-502} by (a) compacting the drawing in the vertical direction, and (b) drawing the path transitive edges that were not drawn in~\cite{JGAA-502}. This approach naturally splits the edges of $G$ into three categories, \emph{path edges}, \emph{cross edges}, and \emph{path transitive edges}. This clearly adds new possibilities to the understanding of the user and allows a system to show the different edge categories separately, without altering the user's mental map.  

\subsection{Compaction} 
Let $G=(V,E)$ be a DAG with $n$ vertices and $m$ edges.  Following the framework of~\cite{ortali2018algorithms,JGAA-502} the vertices of $V$ are placed in a unique $y$-coordinate, which is specified by a topological sorting. Let $T$ be the list of vertices of $V$ in ascending order based on their $y$-coordinates. We start from the bottom and visit each vertex in $T$ in ascending order. For every vertex $v$ in this order we assign a new $y$-coordinate, $y(v)$, following a simple rule that compacts the height of the drawing: ``If $v$ has no incoming edges then we set its $y(v)$ equal to $0$, else we set $y(v)$ equal to $a+1$, where $a$ is the $highest$ $y$-coordinate of the vertices that have edges incoming into $v$."  

 \begin{figure}[]
 \begin{subfigure}{0.5\textwidth}
  \centering
  \includegraphics[height=13cm,width=0.65\linewidth]{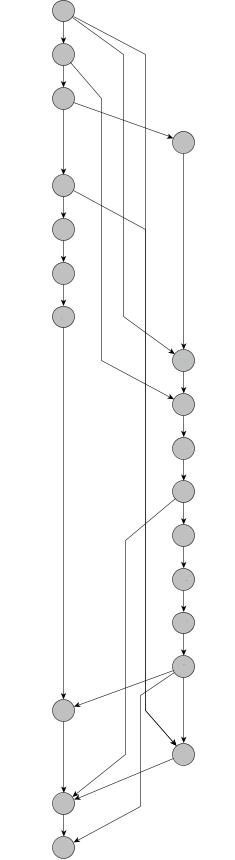}
  \caption{ }
  \label{fig:sfig1}
\end{subfigure}
\begin{subfigure}{.5\textwidth}
  \centering
  \includegraphics[height=13cm,width=0.6\linewidth]{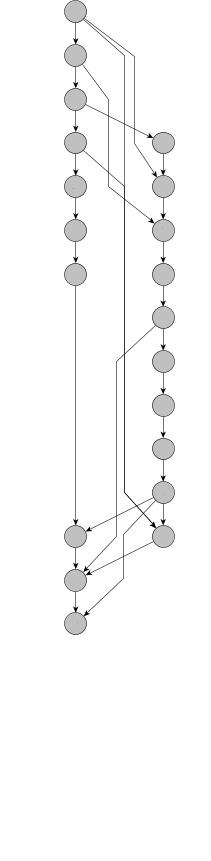}
  \caption{}
  \label{fig:sfig2}
\end{subfigure}
\caption{A DAG $G$ drawn without its path transitive edges: (a) drawing $\Gamma_1$ is computed by Algorithm $PBH$, (b) drawing $\Gamma_2$ is the output after compaction and edge bundling.}
 \label{steps}
\end{figure}

The produced drawings as shown in~Figure~\ref{steps}, have the following simple properties:

\ \\
\textit{Property 1.}\quad Two vertices of the same path are assigned distinct $y$-coordinates.
\ \\
\textit{Property 2.}\quad For every vertex $v$ with $y(v)\neq 0$, there is an incoming edge into $v$ that starts from a vertex $w$  such that $y(v) = y(w) + 1$.

Based on the properties above the height of the compacted drawing of the graph $G$ is at most $L$ and it can be computed in $O(n+m)$ time.

\subsection{Drawing and Bundling the Path Transitive Edges}
An important aspect of our work is the preservation of the mental map of the user that can be expressed, in part, by the reachability information of a DAG.  Recall that path transitive edges are not drawn by the framework of~\cite{ortali2018algorithms,JGAA-502}. In this subsection we show how to bundle and draw these edges while preserving the user's mental map of the previous drawing.  Additionally, one may interact with the drawings by hiding the path transitive edges at the click of a button without changing the user's mental map of the complete drawing.  We describe an algorithm that bundles and draws the path transitive edges using the minimum extra width (minimum extra number of columns) for each (decomposition) path as shown in Figure~\ref{bundle}. The steps of the algorithm are briefly described as follows:


\begin{figure}[]
\begin{subfigure}{.2\textwidth}
  \centering
  \includegraphics[scale=0.45]{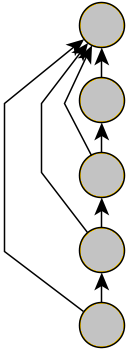}
  \label{fig0:sfig1}
\end{subfigure}
\begin{subfigure}{.2\textwidth}
  \centering
  \includegraphics[scale=0.45]{{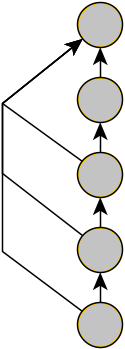}}
  \label{fig1:sfig2}
\end{subfigure}
\hfill
\begin{subfigure}{.2\textwidth}
  \centering
  \includegraphics[scale=0.45]{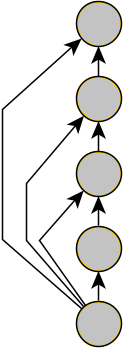}
  \label{fig0:sfig1}
\end{subfigure}
\begin{subfigure}{.2\textwidth}
  \centering
  \includegraphics[scale=0.45]{{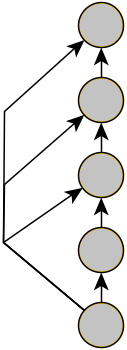}}
  \label{fig1:sfig2}
\end{subfigure}
\caption{Bundling of path transitive edges from left to right: (i) incoming edges into the last vertex of the path, (ii) bundling the incoming edges, (iii) outgoing edges from the first vertex of the path, (iv) bundling the outgoing edges. }
\label{bundle}
\end{figure}

\begin{enumerate}
\item For every vertex of each decomposition path we compute the indegree and outdegree based only on path transitive edges.
\item  If all indegrees and outdegrees are zero the algorithm is over, if not, we select a vertex $v$ with the highest indegree or outdegree and we bundle all the incoming or outgoing edges of $v$, respectively. These bundled edges are represented by an $interval$ with starting and finishing points, the lowest and highest $y$-coordinates of the vertices, respectively.
\item Next, we insert each interval on the left side of the path on the first available column such that the interval does not overlap with another interval.
\item   We remove these edges from the set of path transitive edges, update the indegree and outdegree of the vertices and repeat the selection process.
\item The intervals of the rightmost path, are inserted on the right side of the path in order to avoid potential crossings with cross edges. 
\item A final, post-processing step can be applied because some crossings between intervals/bundled edges can be removed by changing the order of the columns containing them.
\end{enumerate}



The above algorithm can be implemented to run in $O(m+n \log n)$ time by handling the updates of the indegrees and outdegrees carefully, and placing the appropriate intervals in a (Max Heap) Priority Queue.  As expected, the fact that we draw the path transitive edges increases the number of bends, crossings, and area, with respect to not drawing them.

For each decomposition path, suppose we have a set of $b$ intervals such that each interval $I$ has a start point, $s_I$, and a finish point $f_I$. The starting point is the position of the vertex of the interval with the lowest $y$-coordinate. Similarly, the finish point $f_I$ is the position of the node of the interval with the highest $y$-coordinate.  We follow a greedy approach in order to minimize the width (number of columns) for placing the bundled edges.  The approach is similar to Task Scheduling~\cite{GTbook}, for placing the intervals. It uses the optimum number of columns and runs in $O(b \log b)$ time, for each path with $b$ intervals. 
Since the sum of all $b$'s for all paths in a path decomposition is at most $n$ we conclude that the algorithm runs in $O(n \log n)$ time.  For details and proof of correctness see~\cite{GTbook}.


\subsection{Drawing and Bundling the Cross Edges}
\begin{figure}[]
 \begin{subfigure}{0.5\textwidth}
  \centering
         \includegraphics[height=5cm,width=0.5\linewidth]{{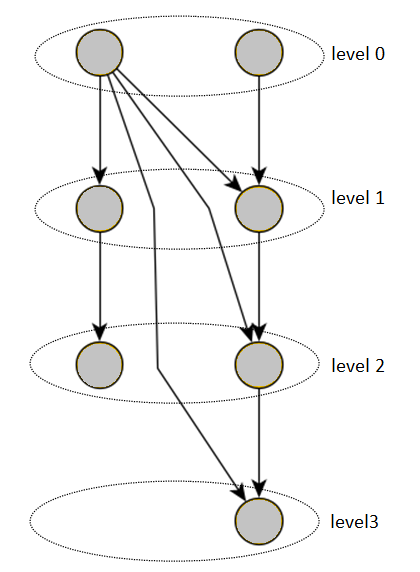}}
         \centering
        \caption{Placing bends on cross edges}
        \label{fig:CrossEdgesBends}
\end{subfigure}
\begin{subfigure}{.5\textwidth}
  \centering
         \includegraphics[height=5cm,width=0.35\linewidth]{{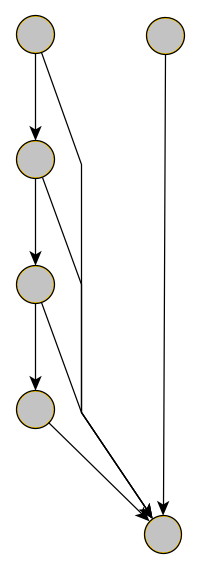}}
         \centering
        \caption{Bundling of cross edges}
        \label{fig:CrossEdgesBundling}
\end{subfigure}
\caption{Examples of bends and bundling of cross edges with a common end node.}
 \label{steps2}
\end{figure}

Cross edges connect vertices that belong to different paths. 
The number of bends of every cross edge depends on the vertical distance of its incident nodes. An example is shown in Figure~\ref{fig:CrossEdgesBends}. 
Each cross edge $(u1,u2)$ has: 
\begin{enumerate}
\item Two bends if the vertical distance between $u1$ and $u2$ is more than two.
\item One bend if the vertical distance between $u1$ and $u2$ is two.
\item The edge is a straight line segment (no bend) if the vertical distance between  $u1$ and $u2$ is one.
\end{enumerate}

We bundle all incoming cross edges for each vertex (except those with one unit of vertical distance from the target). We can place the bundled cross (Figure \ref{fig:CrossEdgesBends}) edges between the paths/channels using the same technique we used for path transitive edges, using a technique that relies on task scheduling as we described above.~Figure \ref{fig:CrossEdgesBundling}  shows an example of bundled cross edges.

\section{Experimental Evaluation and User Study}
In this section we present experimental results obtained by the extended path-based framework and we compare them with the respective experimental results obtained  by running the hierarchical drawing module of OGDF, which is based on the Sugiyama Framework. In order to evaluate the performance, we used the following standard metrics:

\begin{description}
\item[$\bullet$ Number of crossings.]
\item[$\bullet$ Number of bends.]
\item[$\bullet$ Width of the drawing:] The total number of distinct x coordinates that are used by the framework.
\item[$\bullet$ Height of the drawing:]The total number of distinct y coordinates that are used by the framework.  
\item[$\bullet$ Area of the drawing:] The area of the enclosing rectangle.
\end{description}

Based on these metrics, we conducted a number of experiments to evaluate the performance of the two different hierarchical frameworks using a dataset of 20 randomly generated DAGs. 
Additionally, the metrics of PBF could vary depending on the path/channel decomposition algorithm we use and the ordering of the columns. 

\par
\begin{figure}[]
\hspace{-10mm}
 \begin{subfigure}{0.5\textwidth}
 \includegraphics[height=7cm,width=8cm]{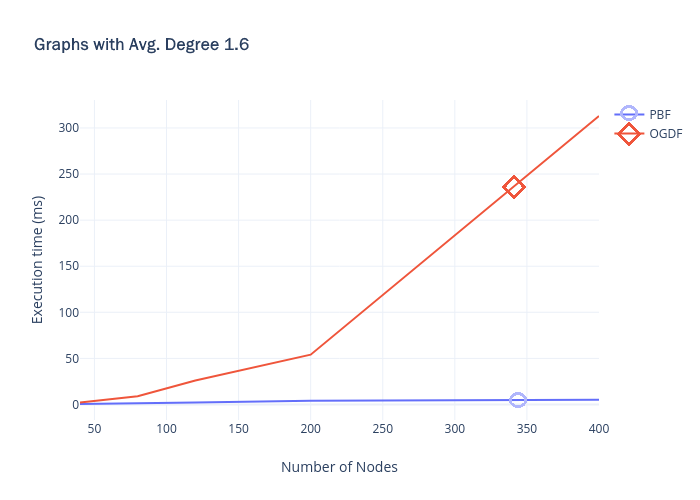}
   \caption{Graphs with average degree 1.6}
   \label{fig:results_pbf}
\end{subfigure}
\begin{subfigure}{.5\textwidth}
  \centering
   \includegraphics[height=7cm,width=7cm]{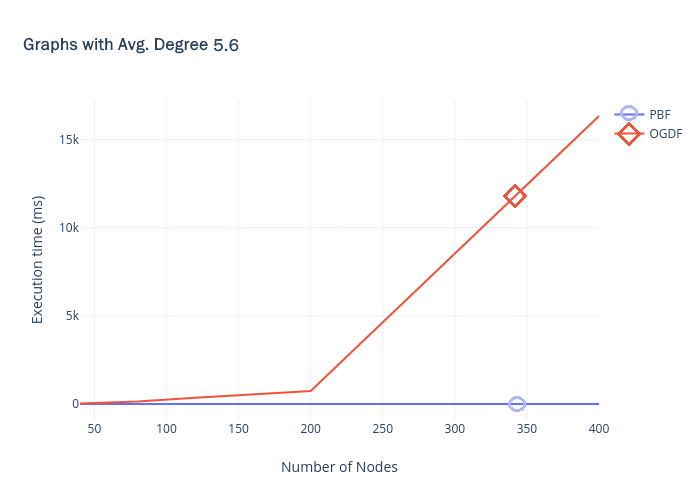}
\caption{Graphs with average degree 5.6}
\label{fig:results_ogdf}
\end{subfigure}
\caption{Execution time of $PBF$ and $OGDF$ on various graphs.}
\label{results_1656}

\end{figure}

In general, our experiments showed that PBF produces readable drawings. Additionally, it clearly partitions the edges into three distinct categories, and vertically aligns certain paths, which may be user/application defined. This may be important in certain applications.
The results showed that our approach differs from the Sugiyama Framework completely, since it examines the graph vertically. The extended PBF performs bundling very efficiently and computes the optimal height of the graph. In most cases, the drawings based on PBF need less area than OGDF and contain fewer bends. On the other hand, OGDF generally has fewer crossings than PBF. This is expected since OGDF places a major computational effort into the crossing minimization step, whereas PBF does not perform any crossing minimization.  Figure~\ref{results_1656} shows that the time for PBF grows linearly in contrast to ODFG where it's time complexity seems to be cubic.
For all the reasons described above this approach seems to be an interesting alternative to the Sugiyama style hierarchical drawings.

\par
Additionally, we conducted another series of experiments in order to validate this statement further. To this respect, we used the  benchmarks found at \url{www.graphdrawing.org}. The archive consists of graphs with 10 to 100 nodes with average degree about 1.6. 
The results of these additional experiments are similar to the results reported above and highlight the fact that the two frameworks focus on different aspects of the graphs and produce vastly different drawings. The new results reinforce our initial consideration that comparing quantitative metrics alone does not lead to a concrete conclusion. Hence, we decided to perform a user study in order to evaluate the readability and clarity of PBF comparing it with the Sugiyama framework from the perspective of a user.

\subsection{User Study}
\noindent\textbf{Users.} We recruited 72 participants. In order to have more accurate and sophisticated results, we selected an audience that was familiar with graph theory and graph drawing styles. More specifically, 35\% were software developers and researchers and 65\% were postgraduate and advanced undergraduate students. 
\\ \\
\noindent\textbf{Training.} We created a Google form and we invited all participants to fill it in. Initially, the users were asked to watch a short video used for training: the tutorial gave a short description of the two hierarchical drawing frameworks (the video is available at \url{https://youtu.be/BWHc2xO4jmI}).

\noindent\textbf{Datasets.} We experimented with a dataset of 3 graph categories with different number of nodes (20 nodes, 50 nodes and 100 nodes, i.e., small, medium and large graphs) with average degree around 1.6 (Table~\ref{tab:graphs_stats}).

\begin{table}[h!]
\centering
\begin{tabular}{c c c c c|} 
 \hline
 Graph & Number of nodes & Number of edges  &Graph Sizes \\
 \hline\hline
    Graph1 & 20 & 31 & Small Graphs \\
    Graph2 & 20 & 31\\
    \hline
    Graph3 & 50 & 82 & Medium Graphs\\
    Graph4 & 50 & 79\\
    \hline
    Graph5 & 100 & 163 & Large Graphs\\
    Graph6 & 100 & 169\\
 \hline
\end{tabular}
  \caption{Graphs dataset.}
  \label{tab:graphs_stats}
\end{table}

\noindent\textbf{Tasks.} We asked the users to answer a set of questions for the two different drawings and carry out a sequence of basic tasks.
Similar to previous user studies (see,
e.g., \cite{di2013exploring},\cite{ghoniem2004comparison},\cite{purchase2012usability}, \cite{DBLP:journals/cgf/DidimoKMT18}), we decided to choose tasks involving graph reading which are easily understandable also to non-expert users. Moreover, we also took into account that the purpose is to evaluate hierarchical drawings and as expected some tasks such as counting incoming or outgoing edges are rather simple and they would not produce useful insights. Thus, we considered the tasks shown in Table~\ref{tab:tasks}.

\begin{table}[ht]
    \centering
    \begin{tabular}{p{0.1\linewidth} | p{0.9\linewidth}}
      \textit{ID} & \textit{Task Description}\\ \hline
1 & Is there a path between the two highlighted vertices? \\ 
2 & How long is the shortest path between the two highlighted vertices? \\ 
3 & Is there a path of length at most 3 that connects the two highlighted nodes?  \\ 
4 & Are all of the green vertices successors of the red vertex? \\ \hline
    \end{tabular}
    \caption{The set of tasks participants had to answer for each of the 2 different graph drawing frameworks over various graphs.}
    \label{tab:tasks}
\end{table}

\begin{figure}[]
\begin{subfigure}{0.5\textwidth}
  \centering
  \includegraphics[scale=0.4]{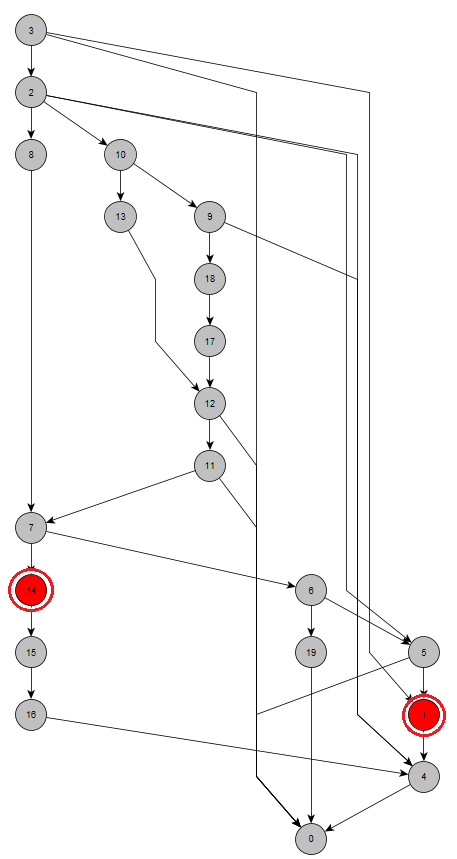}
  \caption{Drawing produced by PBF}
  \label{fig:sfig1}
\end{subfigure}%
\begin{subfigure}{.5\textwidth}
  \centering
  \includegraphics[scale=0.4]{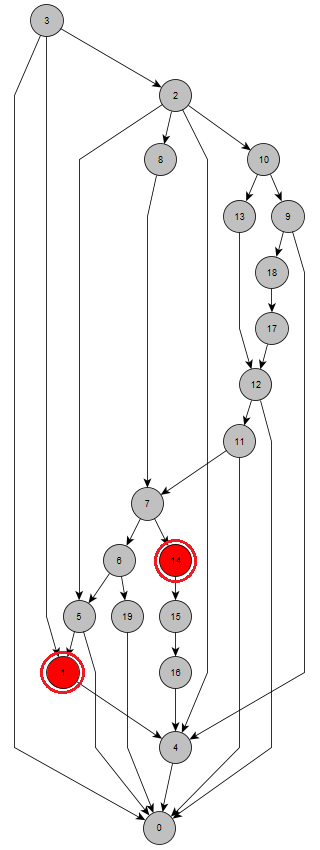}
  \caption{Drawing produced by OGDF}
  \label{fig:sfig2}
\end{subfigure}
\caption{Snapshots of drawings of the same graph used in the user study for the question \textit{"Is there a path between the two highlighted vertices?"}}
\label{is_there_path}
\end{figure}

For questions on Task $2$, the participants had to choose a number as an answer. We do not require numeric answers for all the tasks. More specifically, for Tasks $1$, $3$ and $4$ the participants had to choose an answer among “Yes”, “No”, or “Do Not Know”. There was no time limit, although participants were expected to answer “Do Not Know” if a question was too difficult or too time-consuming to answer. Each of the previous tasks was repeated for each drawing framework model 4 times: i.e., 2 using small size graphs and 2 using medium size graphs. Note that for each question of the same task we used different highlighted nodes.
In total, the number of tasks was 32 i.e., 16 questions for PBF drawings and 16 for OGDF drawings. The questions on small graphs  (20 nodes) preceded those on medium graphs (50 nodes). Finally, to counteract the learning effect, the questions appeared in a randomized order. Figure~\ref{is_there_path} shows a snapshot for the question ``Is there a path between the two highlighted vertices?" for both drawings.

\begin{figure*}[h]
\centering
\hspace{-5mm}
\includegraphics[scale=0.55]{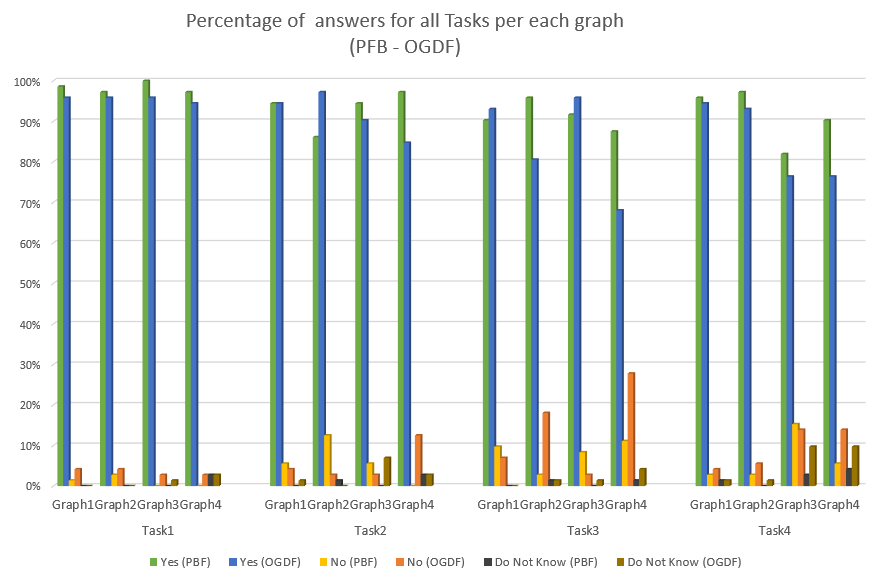}
\caption{Results \textit{(the ratio of participants that answered ``Yes", ``No" and ``Do Not Know" over the total number of answers)}  on the various tasks for each of the drawing framework \textit{(PBF)}, \textit{(OGDF)} over different  graphs.}
\label{tab:results_donot}
\end{figure*}


\begin{figure}[]
\centering
\begin{subfigure}[b]{0.9\textwidth}
\hspace{4mm}
   \includegraphics[height=3.8cm,width=0.9\linewidth]{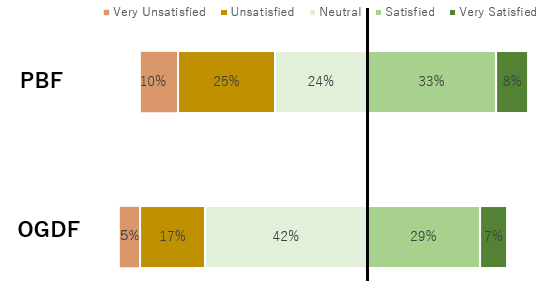}
   \caption{\textit{Percentage results for PBF and OGDF for the question $I$, over $graph5$}}
   \label{} 
\end{subfigure}
\begin{subfigure}[b]{0.9\textwidth}
\centering
   \includegraphics[height=3.8cm,width=0.9\linewidth]{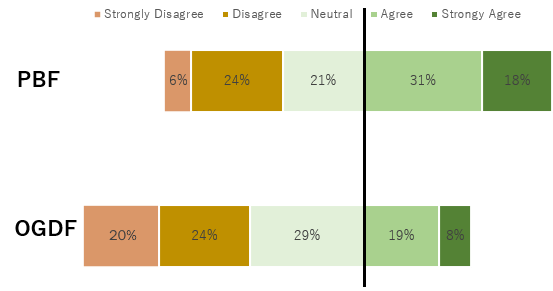}
\caption{Percentage results for PBF and OGDF for the question $II$, over $graph6$}
 \label{}
\end{subfigure}
\caption{\textit{Results for questions I and II.}}
\label{vb}
\end{figure}

\begin{figure}[]
\includegraphics[width=1\linewidth]{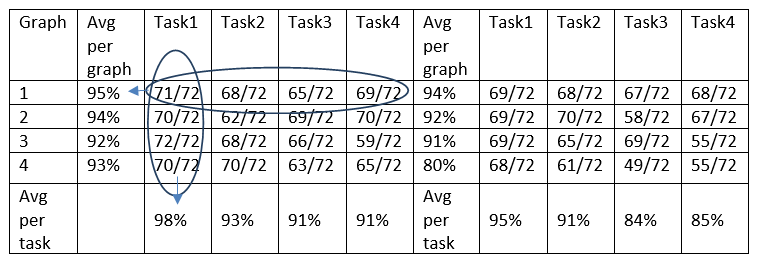}
\caption{Results \textit{(number of correct answers)}  on the various tasks for each of the drawing framework \textit{(PBF)}, \textit{(OGDF)} over different  graphs.}
\label{tab:results}
\end{figure}

\begin{figure*}[]
\centering
\includegraphics[height=4cm,width=1\linewidth]{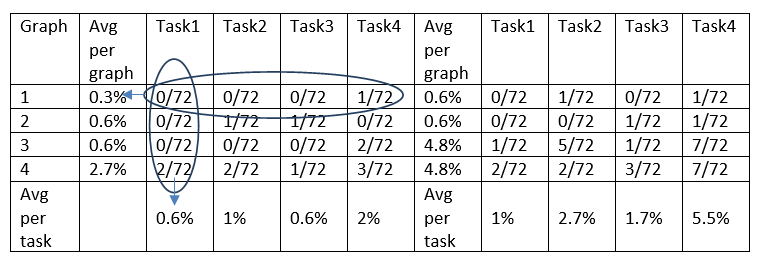}
\caption{Results \textit{(number of ``Do Not Know" answers over the total number of all answers)} on the various tasks for each of the drawing framework \textit{(PBF)}, \textit{(OGDF)} over different  graphs.}
\label{tab:tab_results_donot}
\end{figure*}

\noindent\textbf{Task Based Comparison.} 
For the results, we recorded the total \textit{number of correct answers} for each question of the 2 different drawing frameworks, for all participants. Also note that the “Do Not Know” answer was considered incorrect. More specifically, regarding the first graph (i.e., $graph1$) for all tasks, we have that the users had the same performance in terms of correct answers for both $PBF$ and $OGDF$ drawings. By examining the rest of the results the average percentage revealed that for all the graphs the performance of both drawing frameworks is not significantly different, although it is slightly better for $PBF$ drawings for almost all tasks and drawings. We observe the same when comparing the average percentage for each task: the numbers for PBF drawings are consistently better than the numbers for OGDF drawings, but the differences are small. 
We show a comprehensive visualization of these results in Figure~\ref{tab:results_donot}.  
It shows the ratio of participants that answered  \textit{”Yes”}, \textit{”No”} and \textit{”Do Not Know”} on the various tasks for each of the two drawing frameworks, over the different graphs where we observe again that $PBF$ is slightly better than $OGDF$ for all cases (Tables of Figures~\ref{tab:tab_results_donot},~\ref{tab:results}).
We observe that although the numbers are very low for both, the number of users that answered \textit{”Do Not Know”} for OGDF is often double the corresponding number for PBF, which may imply that some drawings may be more confusing to some users. 

In general, the performance of the participants is slightly better when they are working with PBF drawings than with OGDF drawings.  Since the differences are rather small we cannot extract a concrete conclusion as to which is better for the users. However, it has become clear that the path-based framework is an interesting alternative to the Sugiyama Framework for visualizing hierarchical graphs.  Furthermore, for specific applications, that require to visualize specific paths (such as critical paths) it would be the preferred choice since the nodes of each path are placed on the same $x$-coordinate.

\ \\
\noindent\textbf{Direct Comparison of the two frameworks.}
As a second-level of analysis, we perform a direct comparison of the two drawing frameworks. We used the two (large) graphs of 100 nodes. To this respect we asked the participants to rate each of the 2 models by answering the following questions:
\ \\
\begin{enumerate}[label=\Roman*]
    \item On a scale of 1 to 5, how satisfied are you with the following graph drawings?
    \item Do you believe it would be easy to answer the previous tasks for the following graph?
    \item Which of the following drawings of the same graph do you prefer to use in order to answer the previous tasks?
\end{enumerate}
 Similar to the previous experiments, we had two PBF drawings and two OGDF drawings. Since the objective of this section was to evaluate the usability of both drawing frameworks, using the System Usability Score (SUS) \cite{bangor2009determining}, we asked the users to answer questions $I$ and $II$, by giving a rate using the following scale \textit{$<$Very Unsatisfied, Unsatisfied, Neutral, Satisfied, Very Satisfied$>$} and  \textit{$<$Strongly Disagree, Disagree, Neutral, Agree, Strongly Agree$>$} respectively. The results show that for Question $I$, 41\% 
 of the participants rated $PBF$ from scale 4 and 5, in contrast to 36\%
for $OGDF$. Notice that the answers scale 1 and 2 are worse for PBF.
This probably signifies that the users are not familiar with this new hierarchical drawing style. Question $II$, almost 50\% of the participants rated $PBF$ from scale 4 to 5, in contrast to less than 30\% for $OGDF$, see Figure~\ref{vb}.
\par
\begin{figure}
\centering
\begin{subfigure}[b]{0.95\textwidth}
   \includegraphics[height=12cm,width=0.7\linewidth]{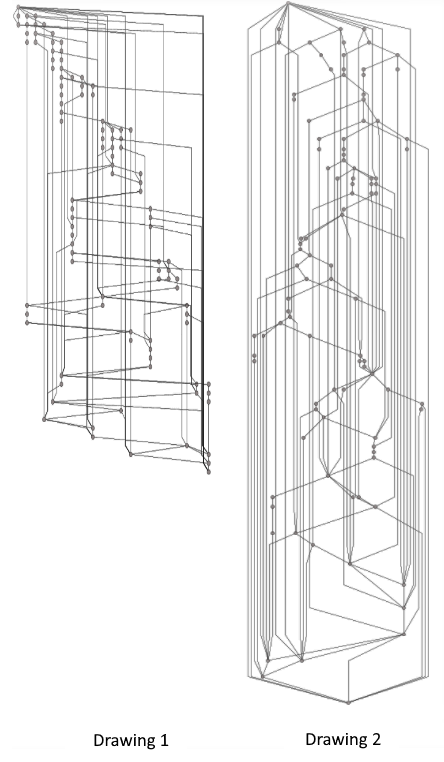}
   \caption{}
   \label{vs_comparison} 
\end{subfigure}
\begin{subfigure}[b]{0.95\textwidth}
\centering
   \includegraphics[height=4cm,width=0.4\linewidth]{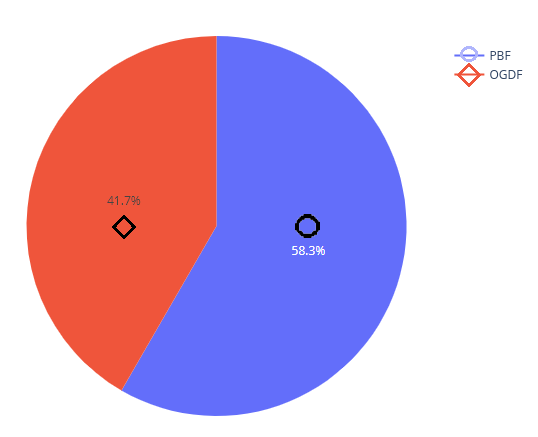}
   \caption{}
   \label{fig:pie_vs}
\end{subfigure}
\caption{In (a) we show snapshots of the same graph as used in our survey. Drawing 1 is the one computed by $PBF$ and Drawing 2 is the one as produced by $OGDF$. In (b) we see the percentage results for the task \textit{"Which of the following drawings of the same graph do you prefer to use in order to answer the previous tasks".}}
\end{figure}

At the end of this user study, we asked the participants to perform a direct comparison of the two drawing frameworks for the same graph, by answering this question: ``Which of the following drawings of the same graph do you prefer to use in order to answer the previous tasks?" (Figure~\ref{vs_comparison}). The results as shown in Figure~\ref{fig:pie_vs} highlight that 58.3\% of the participants stated that they prefer the drawing produced by $PBF$ over the $OGDF$. In terms of statistical significance, the exact (Clopper-Pearson) 95\% Confidence Interval (CI) is (48\%, 72\%) indicating that PBF drawings are preferred by the users over OGDF drawings.  However, given the small differences, we conclude 
that $PBF$ is a significant alternative to the Sugiyama Framework for visualizing hierarchical graphs.

\section{Conclusions}

We present a detailed general-purpose hierarchical graph drawing framework  that is based on the Path Based Framework (PBF)~\cite{JGAA-502}. 
We apply extensive edge bundling to draw all the path transitive edges, and cross edges of the graph and we minimize its height by using compaction. 
%
The experiments revealed that our implementation runs very fast and produces drawings that are readable and efficient. We also evaluated the usability of this new framework compared to $OGDF$ which follows the Sugiyama Framework. The experimental results show that the two frameworks differ considerably. Generally, the drawings produced by our algorithms have lower number of bends and are significantly smaller in area than the ones produced by OGDF, but they have more crossings for sparse graphs. Thus the new approach offers an interesting alternative for visualizing hierarchical graphs, since it focuses on showing important aspects of a graph such as critical paths, path transitive edges, and cross edges. For this reason, this framework may be particularly useful in graph visualization systems that encourage user interaction. Moreover, the 
user evaluation shows that the performance of the participants is slightly better in PBF drawings than in OGDF drawings and the participants prefer PBF in overall rating compared to OGDF.
\bibliographystyle{splncs04}
\clearpage
\bibliography{diffBibBnodesConf} 
\clearpage
\end{document}